%% file: document.tex
\documentclass{llncs}
\pdfoutput=1
\usepackage{graphicx, array}
\usepackage{makeidx}  % allows for indexgeneration
\usepackage[usenames,dvipsnames]{color}
\usepackage{subfig}
\usepackage{amsmath}
\usepackage{amssymb}
\usepackage{url}
\usepackage{clrscode3e}
\usepackage{multirow}
\usepackage[autolanguage]{numprint}
\usepackage{comment}
\usepackage{array}

\newcounter{TodoCounter}

\newcounter{QstCounter}

\newcounter{jdNoteCounter}

\newcounter{mjNoteCounter}

\newcounter{mcNoteCounter}

\newcommand{\MA}{M$_{\text{A}}$}
\newcommand{\MB}{M$_{\text{B}}$}
\newcommand{\MAT}{M$'_{\text{A}}$}

\newcolumntype{L}[1]{>{\raggedright\arraybackslash}p{#1}}
\newcolumntype{C}[1]{>{\centering\arraybackslash}p{#1}}
\newcolumntype{R}[1]{>{\raggedleft\arraybackslash}p{#1}}

\begin{document}
\hyphenation{de-li-ve-ry ana-lysis pa-ra-digm es-tab-lished sce-na-rios acce-le-ra-te va-li-da-tion
mo-del-ling avai-lable go-ver-ning si-mu-la-ted in-stan-tia-ted ex-pe-ri-ments pac-ka-ge re-le-vant si-mu-la-tion pro-duc-ti-vi-ty fle-xi-bi-li-ty rou-ting mo-dels rea-lis-tic me-cha-nism met-hods au-to-ma-ti-cal-ly con-tai-ning im-pro-ve-ments on-de-mand in-he-ren-tly ope-ra-tio-nal com-pri-sing spe-ci-fi-cally dri-ver match-ma-king si-mu-la-tor de-ve-lop-ment ge-ne-rally VALFRAM dia-ries app-li-ca-tion gra-du-al}

\frontmatter          % for the preliminaries
\pagestyle{headings}  % switches on printing of running heads
\mainmatter              % start of the contributions

\title{Data Driven Validation Framework for Multi-agent Activity-based Models}
\titlerunning{Validation Framework}  % abbreviated title (for running head and TOC)

\author{Jan Drchal \and Michal \v{C}ertick\'{y} \and Michal Jakob}
\authorrunning{Drchal, \v{C}ertick\'{y}, Jakob} % abbreviated author list (for running head)


\urldef{\mailsa}\path|drchajan@fel.cvut.cz|
\urldef{\mailsb}\path|{certicky,jakob}@agents.fel.cvut.cz|

\institute{Faculty of Electrical Engineering\\
Czech Technical University in Prague\\
\mailsa\\ \mailsb
}
\maketitle              % typeset the title of the contribution
%\mjnote{Vyhodnocujeme 'validity of models' nebo 'accuracy of models', pricemz model je validni, je-li jeho accuracy dost vysoka? Kdyz tak konzistetne upravit v celem textu.}
\begin{abstract}
Activity-based models, as a specific instance of agent-based models, deal with agents that structure their activity in terms of (daily) activity schedules. An activity schedule consists of a sequence of activity instances, each with its assigned start time, duration and location, together with transport modes used for travel between subsequent activity locations. A critical step in the development of simulation models is validation. Despite the growing importance of activity-based models in modelling transport and mobility, there has been so far no work focusing specifically on statistical validation of such models. 
%Activity-based models are widely used in transport research to generate and study travel demand. 
In this paper, we propose a six-step \emph{Validation Framework for Activity-based Models (VALFRAM)} that allows exploiting historical real-world data to assess the validity of activity-based models. The framework compares temporal and spatial properties and the structure of activity schedules 
%(e.g., activity durations, trip distribution in space or typical activity sequences) 
against real-world travel diaries and origin-destination matrices. We confirm the usefulness of the framework on three real-world activity-based transport models.

% , using a number of metrics to produce a collection of quantitative numeric validity indicators. 

\end{abstract}
%
\input{secIntroduction}
\input{secValidation}

%\input{secDataSources}
\input{secTasks}
\input{secEvaluation}
\input{secConclusion}
\bibliography{references}{}
\bibliographystyle{apalike}

\clearpage
\addtocmark[2]{Author Index} % additional numbered TOC entry
\renewcommand{\indexname}{Author Index}
\printindex
\clearpage
\addtocmark[2]{Subject Index} % additional numbered TOC entry
\markboth{Subject Index}{Subject Index}
\renewcommand{\indexname}{Subject Index}
\end{document}

%% file: secIntroduction.tex
% -*- root: document.tex -*-
\section{Introduction}\label{sec:intro}
%\mjnote{(1) Proc je modelovani (mobility) dulezite\\ }
Transport and mobility have recently become a prominent application area for multi-agent systems and agent-based modelling~\cite{chen2010review}. Models of transport systems offer an objective common ground for discussing policies and compromises \cite{de2011modelling}, help to understand the underlying behaviour of these systems and aid in the actual decision making and transport planning.
%\mjnote{Bylo by dobre podprit referenci ~\cite{schleiffer2002intelligent}}

%\mjnote{(2) proc se jako vhodny pristup jevi agentni aktivitni modely\\ }
Large-scale, complex transport systems, set in various socio-demographic contexts and land-use configurations, are often modelled by simulating the behaviour and interactions of millions of autonomous, self-interested agents. Agent-based modelling paradigm generally provides a high level of detail and allows representing non-linear patterns and phenomena beyond traditional analytical approaches \cite{bonabeau2002agent}. Specific subclass of agent-based models, called {\em activity-based models}, address particularly the need for realistic representation of travel demand and transport-related behaviour.  Unlike traditional trip-based models, activity-based models view travel demand as a consequence of agent's needs to pursue various activities distributed in space and understanding of travel decisions is secondary to a fundamental understanding of activity behaviour \cite{jones1990activity}. 

Gradual methodological shift towards such a behaviourally-oriented modelling paradigm is evident. An early work on the topic is represented by the CARLA model, developed as part of the first comprehensive assessment of behaviourally-oriented approach at Oxford \cite{jones1983understanding}.
%, followed by STARCHILD, which was often referred to as the first operational activity-based model \cite{mcnally1986formation}. 
Later work is represented by the SCHEDULER model -- a cognitive architecture producing activity schedules from long- and short-term calendars and perceptual rules \cite{garling1994computational}, TRANSIMS -- an integrated system of travel forecasting models, including activity scheduler \cite{smith1995transims}, or ALBATROSS -- the first model of complete activity scheduling process automatically estimated from data \cite{arentze2000albatross}.\\
\indent In order to produce dependable and useful results, the model needs to be {\em valid}\footnote{\textit{Valid} model is a model of sufficient \textit{accuracy} (\textit{precision}). We use these terms interchangeably in the following text.} enough. In fact, validity is often considered the most important property of models \cite{klugl2009agent}. The process of quantifying the model validity by determining whether the model is an accurate representation of the studied system is called {\em validation} and the validation process needs to be done thoroughly and throughout all phases of model development \cite{law2009build}.\\
\indent Despite the growing adoption of activity-based models and the generally acknowledged importance of model validation, a validation process for activity-based models in particular has not yet been standardized by a detailed methodological framework. Validation techniques and guidelines are addressed in most modelling textbooks \cite{balci1994validation,law2007simulation} and have even been instantiated in the form of a validation process for general agent-based models \cite{klugl2009agent}; however, such techniques are still too general to provide concrete, practical methodology for the key validation step: statistical validation against real-world data.

In this paper, we address this gap and propose a validation framework entitled VALFRAM (Validation Framework for Activity-based Models), designed specifically for statistically quantifying the validity of {\em activity-based} transport models. The framework relies on the real-world transport behaviour data and quantifies the model validity in terms of clearly defined validation metrics. We illustrate and demonstrate the framework on several activity-based transport models of a real-world region populated by approximately 1 million citizens. 

% Activity schedules generated by modelled agents are matched to real-world data, in form of travel diary surveys and origin-destination matrices, and compared in terms of their temporal, spatial and structural properties. Resulting validity metrics can be used not only to argue the usability of a given model, but also for comparison to alternative models or during the process of model's development.

%% file: secValidation.tex
% -*- root: document.tex -*-
\section{Preliminaries}\label{sec:validation}
%\mjnote{Pokud bude misto, strukturovalbych radeji pomoc nadpisu}
\subsection{Activity-based Models}
%\todo{pridat referenci}
Activity-based models~\cite{ben1996travel} are multiagent models in which the agents plan and execute so-called {\em activity schedules} -- finite sequences of {\em activity instances} interconnected by {\em trips}. Each activity instance needs to have a specific {\em type} (e.g. {\tt work}, {\tt school} or {\tt shop}), {\em location}, desired {\em start time} and {\em duration}. Trips between activity instances are specified by their main
%\footnote{In general, the trips between activities can be multi-modal. In such case, we denote the mode of the longest leg (in terms of travel distance or time) as ``main mode'' and define the whole trip by it.} 
transport mode (e.g. {\tt car} or {\tt public transport}). 

\subsection{Validation Methods}
\vskip 5mm
\noindent
Validation methods in general are usually divided into two types:
\begin{itemize}
  \item {\em Face validation} subsumes all methods that rely on natural human intelligence such as expert assessments of model visualizations. Face validation shows that model's behaviour and outcomes are reasonable and plausible within the frame of the theoretic basis and implicit knowledge of system experts or stake-holders. Face validation is in general incapable of producing quantitative, comparable numeric results. Its basis in implicit expert knowledge and human intelligence also makes it difficult to standardize face validation in a formal methodological framework. In this paper, we therefore focus on statistical validation.
  \item {\em Statistical validation} (sometimes called {\em empirical}) employs statistical measures and tests to compare key properties of the model with the data gathered from the modelled system (usually the original real-world system).   
\end{itemize}

\begin{figure}[!h] 
    \includegraphics[width=1\textwidth]{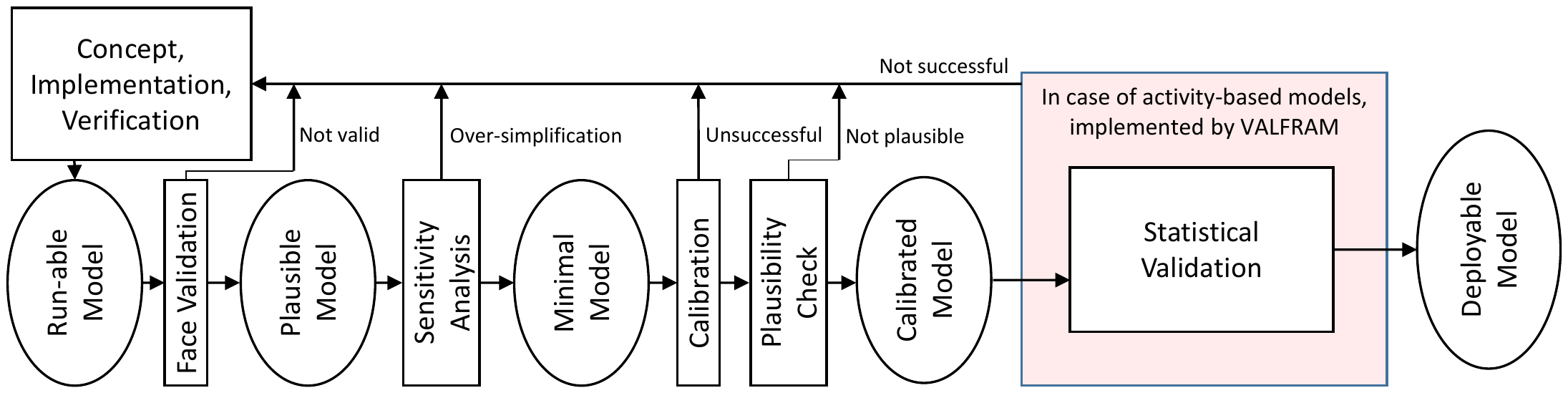}
    \caption{Higher-level validation procedure for agent-based models in general, introduced in \cite{klugl2009agent}. VALFRAM implements the statistical validation step specifically for activity-based models.}
    \label{fig:validation_process_klugl}
  \end{figure}

From a higher-level perspective, VALFRAM can be viewed as an activity-based model-focused implementation of the {\em statistical validation} step of a more comprehensive validation procedure for generic agent-based models, introduced in \cite{klugl2009agent}, as depicted in Figure \ref{fig:validation_process_klugl}. Besides the face and statistical validation, this procedure features other complementary steps such as calibration and sensitivity analysis.

Being set in the context of activity-based modelling, the VALFRAM framework is concerned with the specific properties of activity schedules generated by the agents within the model. These properties are compared to historical real-world data in order to compute a set of numeric similarity metrics. 
%The input real-world data will be described in Section \ref{sec:data} and relevant properties of activity schedules, as well as resulting numeric outputs and methods used to compute them are discussed in detail in Section \ref{sec:tasks}. 

%- on the output for each validation task, we need a number (or few numbers), so that we know when we make model more/less precise.
%  Therefore we need to select (or define) appropriate metrics.
%  There are different metrics used for individual validation tasks (section \ref{sec:tasks}).

%% file: secTasks.tex
% -*- root: document.tex -*-
\section{VALFRAM description}\label{sec:tasks}
In this section a detailed description of VALFRAM is given. We cover validation data, validation objectives and finally measures defined by VALFRAM.
%\mjnote{Nemelo by se to spise jmenovat 'VALFRAM description' - neni to jenom o procesu}
%\mjnote{Pokud bude prostor, bylo by lepe strukturovat podsekcemi, aktualne trochu obtizne citelne}

%\mjnote{**DEL** Zbytek odstavce lze vyhodit, je to sice zajimave, ale ne primo relevantni} It is usually preferable to validate against historical data from real-world systems. However, since such data tends to be difficult or expensive to obtain, models are sometimes validated against data gathered in different models that are known to be valid. This approach, called model alignment, is based on the idea that validity is a transitive relation \cite{axtell1996aligning}.

%\mjnote{Zde by mohla zacinat sekce Data}
\subsection{Data}
A requirement for statistical validation of any model is data capturing the relevant aspects of the behaviour of modelled system, against which the model is validated. To validate an activity-based model, the VALFRAM framework requires two distinct data sets gathered in the modelled system:
\begin{enumerate}
  \item {\em Travel Diaries}: Travel diaries are usually obtained by long-term surveys (taking up to several days), during which participants log all their trips. The resulting data sets contain anonymized information about every participant (usually demographic attributes such as age, gender, etc.), and a collection of all their trips with the following properties: {\em time} and {\em date, duration, transport mode(s)} and {\em purpose} (the activity type at the destination). More detailed travel diaries also contain the {\em locations} of the origin and the destination of each trip.  
  \item {\em Origin-Destination Matrix} (O-D Matrix): The most basic O-D matrices (sometimes called trip tables) are simple two-dimensional square matrices displaying the number of trips travelled between every combination of origin and destination locations during a specified time period (e.g. one day or one hour). The origin and destination locations are usually predefined, mutually exclusive zones covering the area of interest and their size determines the level of detail of the matrix. In real-world systems, O-D matrices may be obtained by roadside monitoring, household surveys or derived from mobile phone networks \cite{caceres2007deriving}.
\end{enumerate}

%\mjnote{**DEL** Nasledujic je zajimave, ale opet ne nutne, protoze mluvi o vyvoji, ne validaci} Note that similar data sets might in some cases be used during the development of the model (usually when machine learning techniques are involved). However, keep in mind that the same data that was used to build the model should never be used for its validation. 

%\mjnote{Zde by mohla zacit sekce 'VALFRAM Validation objectives/properties'}
\subsection{VALFRAM Validation Objectives}
The VALFRAM validation framework is concerned with a couple of specific properties of activity schedules produced by modelled agents. These particular properties need to correlate with the modelled system in order for the model to accurately reproduce the system's transport-related behaviour. At the same time, these properties can actually be validated based on available data sets -- travel diaries and O-D matrices. In particular, we are interested in:
\begin{enumerate}
  \item[A.] {\em Activities} and their:
  	\begin{enumerate}
  	  \item[1.] {\em temporal} properties (start times and durations),
  	  \item[2.] {\em spatial} properties (distribution of activity locations in space),
  	  \item[3.] {\em structure} of activity sequences (typical arrangement of successive activity types).
  	\end{enumerate}
  \item[B.] {\em Trips} and their:
  	\begin{enumerate}
  	  \item[1.] {\em temporal} properties (transport mode choice in different times of day; durations of trips),
  	  \item[2.] {\em spatial} properties (distribution of trip's origin-destination pairs in space),
  	  \item[3.] {\em structure} of transport mode choice (typical mode for each destination activity type).
  	\end{enumerate}
\end{enumerate}

%\mjnote{Zde by mohla zacit sekce 'VALFRAM Validation metrics'}
\subsection{VALFRAM Validation Metrics}
To validate these properties of interest, we need to perform six validation steps (A1, A2, A3, B1, B2, B3), as depicted in Table \ref{tab:steps} and detailed in the rest of this section. In each validation step, we compute specific numeric metrics (statistics). 
%The idea is to have easily computable measures which, if possible, allow for hypothesis testing\footnote{By hypothesis testing we mean goodness-of-fit methods used to compare distributions of various properties of model and validation set. These methods output \mbox{p-values} indicating whether the \textit{null hypothesis} (distributions are the same) should be rejected or not. P-value lower than \textit{significance level} $\alpha$ supports evidence against the \textit{null hypothesis}.}. 
For all metrics, higher values of these statistics indicate a larger difference between the model and validation set, i.e., lower accuracy.

\newcommand{\textAoneMethod}{Compare the distributions of {\em start times} and {\em durations} for each activity type using Kolmogorov-Smirnov (KS) statistic.}
\newcommand{\textAtwoMethod}{Compare distribution of each activity type in 2D space using RMSE. Plot heat maps for additional feedback.}
\newcommand{\textAthreeMethod}{{\em i)} Compare activity counts within activity schedules using $\chi^2$ statistics. {\em ii)} Compare distributions of activity schedule subsequences  as n-grams profiles using $\chi^2$ statistics.}
\newcommand{\textAoneDataset}{Travel Diaries}
\newcommand{\textAtwoDataset}{Space-aware Travel Diaries}
\newcommand{\textAthreeDataset}{Travel Diaries}
\newcommand{\textBoneMethod}{Compare the distribution of selected {\em modes} by {\em time of day} and the distribution of {\em travel times} by {\em mode} using $\chi^2$ and KS statistics. }
\newcommand{\textBtwoMethod}{Compute the distance between generated and real-world O-D matrix using RMSE.}
\newcommand{\textBthreeMethod}{Compare the distribution of selected {\em transport mode} for each type of {\em target activity} type using $\chi^2$ statistics.}
\newcommand{\textBoneDataset}{Travel Diaries}
\newcommand{\textBtwoDataset}{Origin-Destination Matrix}
\newcommand{\textBthreeDataset}{Travel Diaries}

\setlength{\extrarowheight}{2pt}
\setlength{\arraycolsep}{2pt}
\vspace{-5mm}
\begin{table}[h]
\scriptsize
$
\begin{array}{m{0.1cm} | m{3.4cm} | m{1.7cm} | m{3.4cm}| m{1.7cm} |}
\cline{2-5}
 & \multicolumn{2}{c|}{\large {\bf A. Activities}} & \multicolumn{2}{c|}{\large {\bf B. Trips}} \\  \cline{2-5} 
 & {\scriptsize \hfill Task} & {\scriptsize \hfill Data set} & {\scriptsize \hfill Task} & {\scriptsize \hfill Data set}        \\[-0.4mm] \hline
\multicolumn{1}{|l|}{{\bf \hfill 1. Time} } & \textAoneMethod & \textAoneDataset & \textBoneMethod  & \textBoneDataset \\ \hline
\multicolumn{1}{|l|}{{\bf \hfill 2. Space} } & \textAtwoMethod & \textAtwoDataset & \textBtwoMethod  & \textBtwoDataset \\ \hline
\multicolumn{1}{|l|}{{\bf \hfill 3. Structure} } & \textAthreeMethod & \textAthreeDataset & \textBthreeMethod  & \textBthreeDataset \\ \hline 
\end{array} 
$
\vskip 0.4cm
\caption{Six validation steps of VALFRAM framework and corresponding validation data sets needed for each of them.\label{tab:steps}}
%\mjnote{Nelze v texu udelat nejaky padding - v te tabulce jo to hodne namackane - poladime v camera-ready HD}
\end{table}
\vspace{-10mm}
\input{subSecA1}

\input{subSecA2}

\input{subSecA3}
\input{subSecB1}
\input{subSecB2}
\input{subSecB3}

%% file: subSecA1.tex
\vspace{2mm} \noindent {\bf A1. Activities in Time:}
The comparison of activity distributions in time is realized by means of a well-established Kolmogorov-Smirnov two-sample statistic~\cite{Hollander2013}. VALFRAM applies the method to start time distributions $p(\text{start}|\text{act. type})$ as well as to duration distributions $p(\text{duration}|\text{act. type})$.
%\footnote{We have also considered different methods comparing continuous univariate probability distributions including Anderson--Darling and Cram\'{e}r--von Mises statistics.}

The statistic is defined as the maximum deviation between the empirical cumulative distribution functions $F_M$ and $F_V$ which are based on the model and validation data distributions: $d_{KS} = \sup_{x}|F_M(x)-F_V(x)|$. The values lie in the interval $[0, 1]$.

\begin{figure}[!ht]
    \subfloat[\texttt{work} activity start time]{%
      \label{fig:a1_example}
      \includegraphics[height=4.0cm]{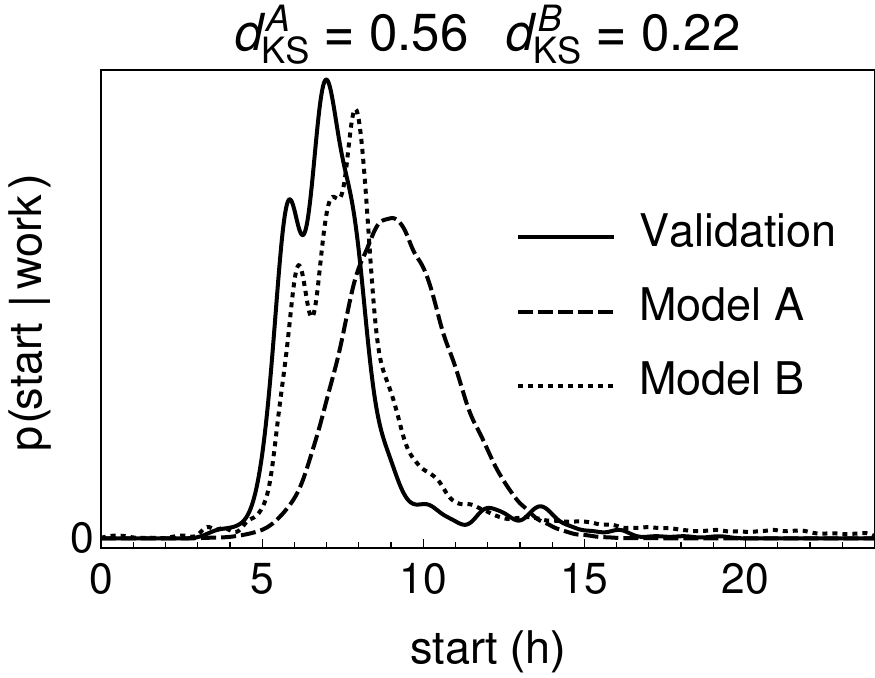}
    }
    \hfill
    \subfloat[Modeled area, \texttt{sleep} activity]{%
      \label{fig:a2_map_heat}
      \includegraphics[height=3.8cm]{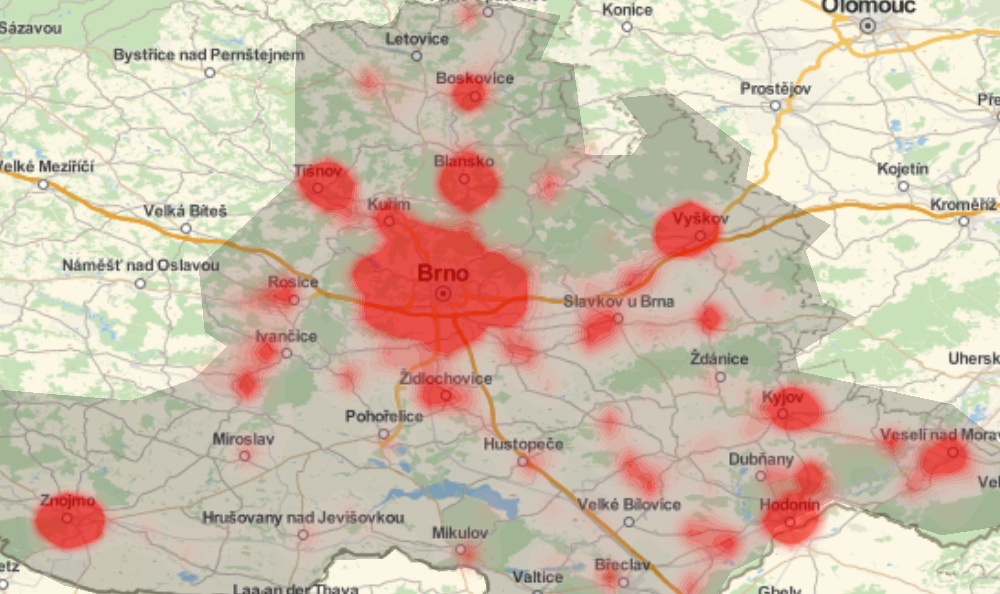}
    }
    \caption{Start time distributions for \texttt{work} activity shown for validation data and two different models (a) including Kolmogorov-Smirnov statistics. Modelled area including \texttt{sleep} activity spatial PDF visualized as a heat map (b). }
  \vspace{-5mm}
  \end{figure}

Figure~\ref{fig:a1_example} shows an example application of the Kolmogorov-Smirnov statistic comparing two different models to validation data.  

%% file: subSecA2.tex
% -*- root: document.tex -*-
\medskip
\noindent {\bf A2. Activities in Space:}
The comparison of activity distributions in space is performed separately for every activity type. Unlike in the previous step, the distributions are two-dimensional (latitude, longitude or projected coordinates). The process consists of the following steps. First, bivariate empirical cumulative distribution functions (ECDFs) $F_M$ and $F_V$ are constructed using coordinate data for both model and validation data, respectively. Second, $F_M$ and $F_V$ are regularly sampled getting matrices $E^M$ and $E^V$ both having $m$ rows and $n$ columns. Third, Root Mean Squared Error (RMSE) of the two matrices is computed using $d_{ecdf} = \sqrt{\sum_{i=1}^m\sum_{j=1}^n\left(E^M_{ij}-E^V_{ij}\right)^2/(m n)}$. As $E^M_{ij}\leq 1$ and $E^V_{ij}\leq 1$, the measure $d_{ecdf}$ is again limited to the $[0,1]$ interval. 

Figure~\ref{fig:a2_map_heat} shows the spatial probability distribution function (PDF) of \texttt{sleep} type activities on the validation set visualized as a heat map. The probability distribution was approximated from data using Gaussian kernels. Similar heat maps might be helpful when developing a model as they can show where problems or imprecisions are.

%% file: subSecA3.tex
% -*- root: document.tex -*-
\medskip
\noindent {\bf A3. Structure of Activities:}
In the previous steps, we examined the activity distributions in time and space. 
%\opt{The analysis focused on distributions based on activity type. We were checking these distributions through the whole datasets (either validation or generated by model).} 
In this step, we consider the activity composition of the entire activity schedules. We propose a measure which compares distributions of \textit{activity counts} in activity schedules as well as a measure comparing the distribution of possible \textit{activity type sequences}.

{\em Activity Count:}
The comparison of activity counts in activity schedules is based on a well-known Pearson's chi-square test~\cite{Sokal1994}. The procedure is performed separately for each activity type. First, frequencies $f^M_i$ and $f^V_i$ for the count $i$ are collected for both model and validation data. Validation data frequencies $f^V_i$ are then used to get count proportions $p^V_i$ and in turn validation frequencies $s^V_i$ scaled to match the sum of model's frequencies ($\sum_i{s^V_i} = \sum_i{f^M_i}$). Using $f^M_i$ and $s^V_i$ chi-square statistic is computed as $\chi^2 = \sum_i{\left(f^M_i - s^V_i\right)^2/s^V_i}$.

{\em Activity Sequences:} We also compare activity sequence distributions. The method is based on the well-established text mining techniques \cite{Cavnar1994} \cite{Manning1999}. Particularly, we compare \textit{\mbox{n-gram} profiles} using chi-square statistic.
N-gram is a continuous subsequence of the original sequence having a length exactly $n$. Consider an example activity schedule consisting of the following activity sequence: $\langle \texttt{none}, \texttt{sleep}, \texttt{work}, \texttt{leisure}, \texttt{sleep}, \texttt{none}\rangle$\footnote{Note, that \texttt{none} activities are added to the beginning and end of the activity schedule in order to preserve information about initial/terminal activity.}. The set of all \mbox{2-grams} (bigrams) is then: $\left\{\langle \texttt{none}, \texttt{sleep}\rangle,\langle \texttt{sleep}, \texttt{work}\rangle,\langle \texttt{work}, \texttt{leisure}\rangle,\langle \texttt{leisure}, \texttt{sleep}\rangle,\right.$ $\left.\langle \texttt{sleep}, \texttt{none}\rangle\right\}$.
We create an \textit{n-gram profile} by counting frequencies of all \mbox{n-grams} in a range $n\in\{1,2,\cdots,k\}$ for all activity schedules. All the $N$ \mbox{n-grams} are then sorted by their counts in a decreasing order so that the counts are $f_i \geq f_j$ for any two n-grams $i$ and $j$ where $1\leq i<j\leq N$ (for a tie $f_i = f_j$ one should sort in the lexicographical order). We only work with a proportion $P$ of n-grams having the highest count in the profile. More precisely, we take only the first $M$ \mbox{n-grams}, where $M$ is the highest value for which $\sum_{i=1}^M{f_i} \leq P\sum_{i=1}^N{f_i}$ is true.

In order to compare n-gram profiles of model and validation data, we employ chi-square statistic matching both profiles by the corresponding \mbox{n-grams} (only \mbox{n-grams} found in both profiles are considered). 

%% file: subSecB1.tex
% -*- root: document.tex -*-

\medskip \noindent {\bf B1. Trips in Time:}
The validation of trips in time consists of two sub-steps: a comparison of mode distributions for a given time of day and a comparison of travel time distributions for selected modes. 

{\em Modes by Time of Day:}
The comparison of mode distributions for a given time of day, i.e., $p(\text{mode} | \text{time range})$, is based on exactly the same approach which we used to compare activity counts (validation step A3): the $\chi^2$ statistic is computed for mode frequencies of trips starting in a selected time interval. We suggest computing $\chi^2$ statistic for twenty four one-hour intervals per day, although other partitionings are possible.

{\em Travel Times per Mode:}
Travel time distributions for modes $p(\text{travel time}|\text{mode})$ are validated in the same way as activities in time (see validation step A1) using Kolmogorov-Smirnov statistic $d_{KS}$. 

%% file: subSecB2.tex
% -*- root: document.tex -*-

\vspace{2mm}\noindent {\bf B2. Trips in Space:}
In order to validate trip distributions in space, we propose a symmetrical dissimilarity measure based on O-D matrix comparison. The algorithm is realized in three consecutive steps. First, O-D matrices are rearranged to use a common set of origins and destinations. Second, both matrices are scaled to make trip counts comparable. Third, RMSE for all elements which have non zero trip count in either of the matrices is computed.

The algorithm starts with two O-D matrices: model matrix $M$ and validation matrix $V$. Each element $M_{ij}$ (or $V_{ij}$) represents a count of trips between origin $i$ and destination $j$. The positional information (i.e., latitude/longitude or other type of coordinates) is denoted $m_i, m_j \in C_M$  for model and similarly $v_i, v_j \in C_V$ for validation data where $C_M$ and $C_V$ are sets of all possible coordinates (e.g., all traffic network nodes). 

Note that in most practical cases $C_M \neq C_V$. As an example we can have precise GPS coordinates generated by the model, however, only approximate or aggregated trip locations from validation travel diaries. As we have to work with the same locations in order to compare the O-D matrices, we need to select a common set of coordinates $C$. In practice, this would be typically the validation data location set ($C = C_V$) while all locations from $C_M$ must be projected to it by replacing each $m_i$ by its closest counterpart in $C$. This might eventually lead to resizing of the O-D matrix $M$ as more origins/destinations might get aggregated into a single row/column.

In many cases the total number of trips in $M$ and $V$ can be vastly different. The second step of the algorithm scales both $M$ and $V$ to a total element sum of one:  $M_{ij}' =\frac{M_{ij}}{\sum_{i}\sum_{j}M_{ij}}$ and $V_{ij}' =\frac{V_{ij}}{\sum_{i}\sum_{j}V_{ij}}$. Each element of both $M_{ij}' $ and $V_{ij}' $ now represents a relative traffic volume between origin $i$ and destination $j$.

Finally, we compute the O-D matrix distance using the following equation:
\begin{equation}
d_{OD} = \sqrt{\frac{\sum_{i}\sum_{j}\left(M'_{ij}-V'_{ij}\right)^2}{\left|\left\{(i,j):M'_{ij}>0 \lor V'_{ij}>0\right\}\right|}}.
\end{equation}
Note that the equation is RMSE computed over all origin-destination pairs which appear either in $M_{ij}' $, $V_{ij}' $ or in both. We have decided to ignore the elements which are zero in both matrices as these might represent trips which might not be possible at all (i.e., not connected by the transport network). Possible values of $d_{OD}$ lie in interval $[0, 1]$ (the upper bound is given by $M'_{ij} \leq 1$ and $V'_{ij} \leq 1$).

%% file: subSecB3.tex
% -*- root: document.tex -*-

\medskip \noindent {\bf B3. Mode for Target Activity Type:}
The validation of the mode choice for target activity type is again based on $\chi^2$ statistic. Here, we collect counts per each mode for each target activity of choice.

%% file: secEvaluation.tex
% -*- root: document.tex -*-
\section{VALFRAM Evaluation}\label{sec:evaluation}
%What do we want from validation framework?
%- Correlates with face/expert validation
%- Reinforces it with statistical basis
%- Enables quantification -> and therefore comparison of different models

%So we'll do this:
%1. take 2 models; 
%2. express some hypotheses about them based on our expert insight
%3. validate them using VALFRAM and see how the results correlate with it

%- ako to robime a preco (statisticke potvrdenie expertnej validacie? experti: jeden model je horsi ako druhy. valid. fram. should confirm this. posilnit expertny posudok a kvantifikovat ho. )
% hypotheses
%- ake su vysledky
%- diskusia o vysledkoch

In general, we expect a statistical validation framework to meet three key conditions:
\begin{enumerate}
  \item The framework quantifies the precision of the validated models in a way which allows comparing model's accuracy in replicating different aspects of the beahviour of the modelled system.
  \item Data required for validation are available.
  \item Validation results produced by the framework correlate with the expectations based on expert insight and face validation.
\end{enumerate} 

VALFRAM meets conditions 1 and 2 for activity-based models by explicitly expressing the spatial, temporal and structural properties of activities and trips, using only travel diaries and O-D matrices. To evaluate it with respect to condition 3, we have built three different activity-based models, formulated hypotheses about them based on our expert insight and used VALFRAM to validate both of them. 

\subsection{Evaluation Models}
%\mjnote{Zde by mohla byt sekce 'Framework test/evaluation models' (nerad bych pouzival validation, to by byla validation na druhou, i kdyz by to asi bylo nejpresnejsi)}
The first model, denoted \MA\ (model A), is a rule-based model inspired by ALBATROSS\footnote{Although we call \MA\ the rule-based model, it estimates activity count, durations and occasionally start times using linear-regression models based on data. All other activity schedule properties are based on rules constructed using expert knowledge.}~\cite{arentze2000albatross}. The second model, denoted \MB, is a fully data-driven model based on Recurrent Neural Networks (RNNs). More specifically, the model employs fully-connected Long-Short Term Memory (LSTM) units \cite{hochreiter1997long} and several sets of softmax output units. Given the training dataset based on travel diaries, the model is trained to repetitively take current activity type and its end time as input in order to produce a trip (including trip duration and main mode) and the following activity (defined by type and duration). As \MB\ is currently unable to generate spatial component of the schedules (e.g., activity locations), VALFRAM steps A2 and B2 are evaluated on a predecessor of \MA\ denoted \MAT\ (model A$'$). \MAT uses a less sophisticated approach to select activity locations. 

All \MA, \MB\ and \MAT\ models were used to generate  a sample of $\numprint{100000}$ activity schedules. Our validation set $V$ contained approximately $\numprint{1800}$ schedules. Such a disproportion is typical in reality, since obtaining real-world data tends to be more costly than obtaining synthetic data from model. All the data used in this study cover a single workday. An overview of the modelled area is depicted in Figure~\ref{fig:a2_map_heat}.

In the following text we present five hypotheses based on our insight of models. Note that all VALFRAM steps A1 through B3 are performed in order to evaluate them.\\

\subsection{Test Hypotheses}
% \mjnote{Zde by mohla byt sekce 'Test hypothesis' }
%covers A1
\noindent{\bf Hypothesis 1:} The rule-based model \MA\ uses very simple linear classifier for decisions on activity start times, so it will likely perform worse than the RNN-based model in their assignment. On the other hand, the activity scheduler in \MA\ performs schedule optimization, during which it adapts activity durations according to rules psychologically plausible. This should produce more realistic behaviour than the purely data-driven RNN model\footnote{At least given the limited size of the RNN training dataset.}. 

\begin{figure}[!ht]
    \subfloat[start time]{%
      \label{fig:a1_starts}
      \includegraphics[height=2.3cm]{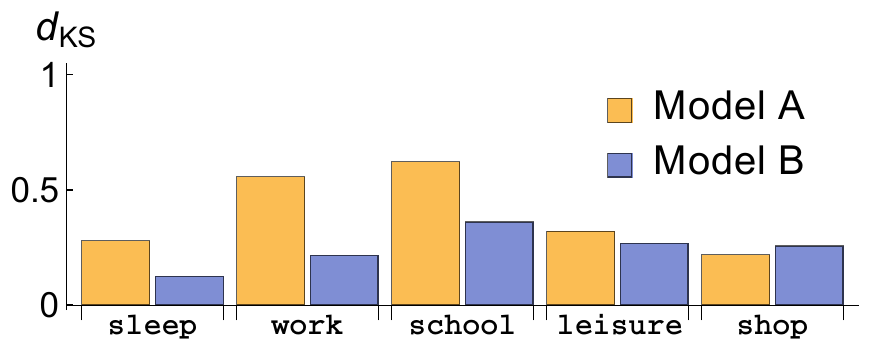}
    }
    \hfill
    \subfloat[duration]{%
      \label{fig:a1_durations}
      \includegraphics[height=2.3cm]{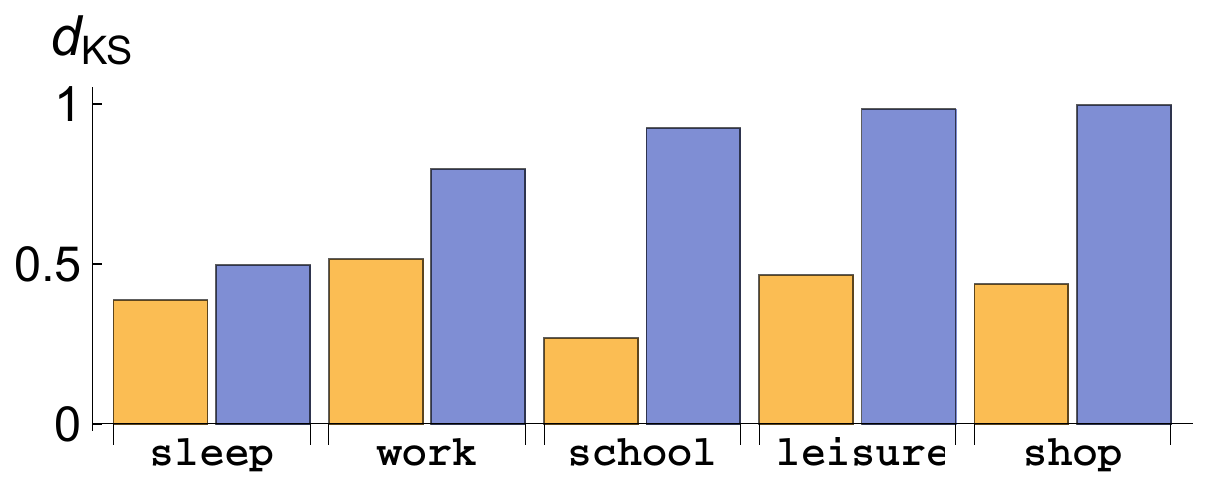}
    }
    \caption{An example of activity in time comparison. The values of $d_{KS}$ are shown for both models \MA\ and \MB. \MB\  outperforms \MA\ on start times while the situation is the opposite for durations. }
    \label{fig:a1_all}
  \end{figure}

Step A1 of VALFRAM confirms the hypothesis. Figure \ref{fig:a1_starts} depicts the distributions $p(\text{start}|\texttt{work})$ for validation data V and models \MA\ and \MB. The values $d_{KS}^A > d_{KS}^B$ indicate the higher precision of the RNN model, with the most significant difference in the case of {\tt work} and {\tt school} activities. On the other hand, Figure~\ref{fig:a1_durations} shows that \MA\  outperforms \MB\ in terms of activity durations.

%\noindent{\em Hypothesis:} Since \MB\ is currently unable to generate spatial components of the schedules (e.g., activity locations), VALFRAM steps A2 and B2 are evaluated on a predecessor of \MA\ denoted \MAT\ (model A$'$). \MAT uses a less sophisticated approach to select activity locations. 

%covers A3
\noindent{\bf Hypothesis 2:} Activity sequences of real-world system tend to be harder to replicate using simple rule-based models than robust data-driven approaches.

Results of the step A3 (\textit{activity counts}) for all the activity types are shown in Table~\ref{tab:a3_activity_counts}. The data-driven model \MB\ outperforms \MA\ with the exception of the \texttt{leisure} activity (which we later found to be insufficiently covered by the RNN training data). Note that both \MA\ and \MB\ give the same $\chi^2$ value for the \texttt{sleep} activity which is caused by the fact that both models generate daily schedules having strictly two \texttt{sleep} activities in the current setup. For the step A3 (\textit{activity sequences}) we got the following results for both models using the proportion $P=0.9$ and $k = 11$ (same as the longest sequence in data): $\chi^2\approx 8.4\times 10^5$ for \MA\ and $\chi^2\approx 2.6\times10^5$ for \MB\ showing superiority of the RNN model. 

\begin{table}[!ht]

    \begin{center}
    \begin{tabular}{ l *{5}{C{1.3cm}} }
    \hline
Model & \texttt{sleep} & \texttt{work} & \texttt{school} & \texttt{leisure} & \texttt{shop}\\\hline
\MA & 21468.1 & 2889.3 & 542.2 & \textbf{1750.3} & 974.2 \\
\MB & 21468.1 & \textbf{255.7} & \textbf{293.8} & 4625.7 & \textbf{773.8} \\

    \hline
    \end{tabular}
   \end{center}
  \caption{Activity counts for selected activities ($\chi^2$ statistic). Model \MB\  
  outperforms model \MA\  with the exception of the \texttt{leisure} activity type.}
  \label{tab:a3_activity_counts}
  \vspace{-10mm}
  \end{table}

%covers A3, B1a
\noindent{\bf Hypothesis 3:} While rule-based model optimizes the whole daily activity plans, RNN-based model works sequentially and schedules new activity based only on the previous ones. Therefore, it will be less precise towards the end of the day.

\begin{figure}[!ht]

    \includegraphics[width=\textwidth]{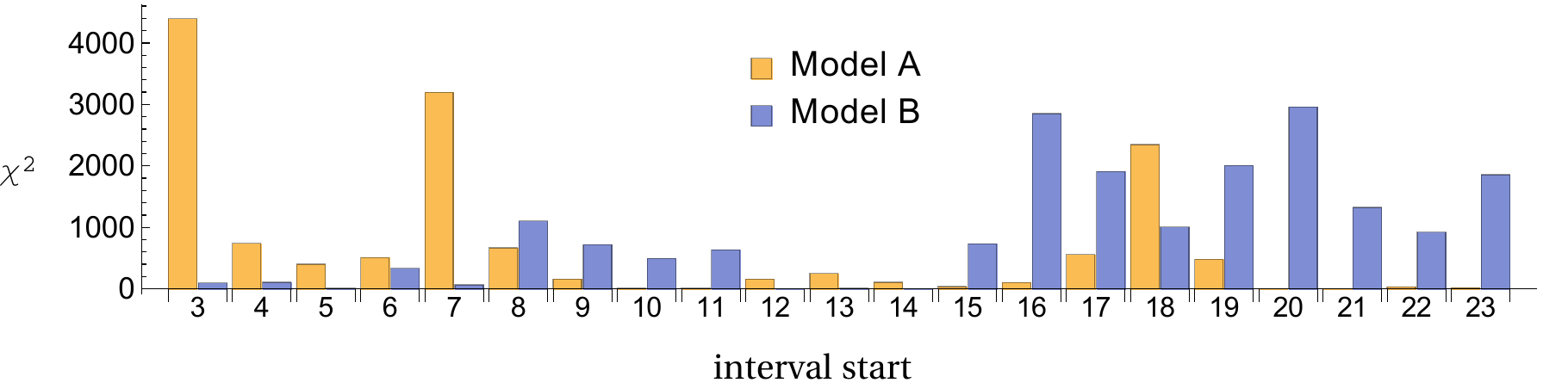}

    \caption{Modes by the time of day. The figure shows a comparison of $\chi^2$ values for \texttt{car} and \texttt{public transport} modes for one hour intervals between 3:00 and 23:00.}
    \label{fig:b1_modes_by_time}
  \end{figure}

By a further analysis of step A3 (\textit{activity sequences}), which involved the comparison of a set of n-grams having highest frequency difference, we have, indeed, found that the RNN model tends to be less precise towards the end of the generated activity sequence resulting in schedules not ended by the \texttt{sleep} activity in a number of cases. Moreover, Figure~\ref{fig:b1_modes_by_time} shows a comparison of \textit{mode by time of day} selection $\chi^2$ values (step B1) for \MA\ and \MB\ showing that although \MB\ is initially more precise it eventually degrades and the rule-based model \MA\ prevails.

%covers B1b, B3
\noindent{\bf Hypothesis 4:} Unlike the rule-based model, the RNN model has no access to trip-planning data (i.e., transport network, timetables) which will decrease its performance in selecting trip modes.

For the step B1 (\textit{travel times per mode}) we got $d^{A}_{KS} = 0.22 < d^{B}_{KS} = 0.31$ for \texttt{car} and $d^{A}_{KS} = 0.37 < d^{B}_{KS} = 0.43$ for \texttt{public transport} modes. Results of the step B3 are summarized in Table~\ref{tab:b3_target_activity_for_mode} also supporting the superiority of \MA\ in modelling mode selection.

\begin{table}[!ht]

    \begin{center}
    \begin{tabular}{ l *{5}{C{1.3cm}} }
    \hline
Model & \texttt{sleep} & \texttt{work} & \texttt{school} & \texttt{leisure} & \texttt{shop}\\\hline
\MA & \textbf{562} & \textbf{1371.7} & \textbf{1120} & 12817.3 & \textbf{5} \\
\MB & 2875.2 & 3437.9 & 7286.2 & \textbf{475.1} & 2507.3 \\

    \hline
    \end{tabular}
   \end{center}
  \caption{Transport mode selection for target activity type ($\chi^2$ statistic). Model \MA\  outperforms model \MB\ in four out of five activity types.}
  \label{tab:b3_target_activity_for_mode}
  \vspace{-7mm}  
\end{table}

%covers A2, B2
\noindent{\bf Hypothesis 5:} Model \MAT\ will be inferior to \MA\ as it uses an oversimplified activity location selection.

\begin{figure}[!ht]
\begin{center}
 \includegraphics[height=2.2cm]{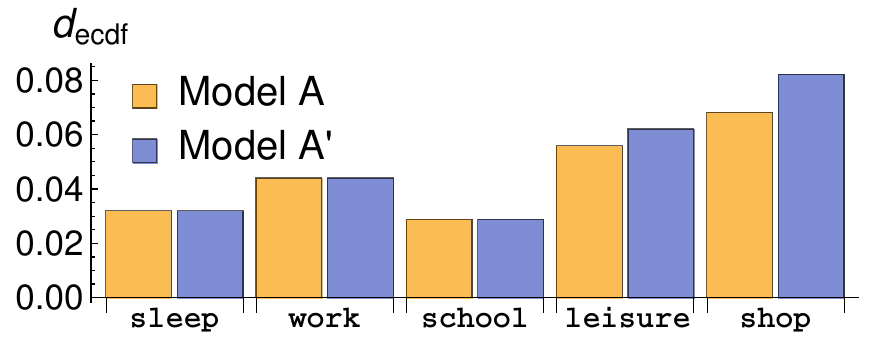}
\end{center}
 \caption{Activities in space: comparison of Model A to Model A$'$. \MAT\ is inferior to \MA\ for flexible activities ($d^A_{ecdf} < d^{A'}_{ecdf}$) based on $18\times 31$ ECDF matrices.}
 \label{fig:fig_a2_ecdf}
   \vspace{-10mm}
\end{figure}

For the step A2 this is clearly demonstrated in Figure~\ref{fig:fig_a2_ecdf} by $d^A_{ecdf} < d^{A'}_{ecdf}$ for the \texttt{leisure} and \texttt{shop} activities (only activity types affected by the algorithm selecting activity locations). For the step B2 we get $d^{A}_{OD} = 3.7\times 10^{-4} < d^{A'}_{OD} = 4.8\times 10^{-4}$ which again supports the hypothesised improvement of A over A$'$.

%% file: secConclusion.tex
% -*- root: document.tex -*-
\section{Conclusion}\label{sec:conclusion}
We have introduced a detailed methodological framework for data-driven {\em statistical validation} of multiagent activity-based transport models. The VALFRAM framework compares activity-based models against real-world {\em travel diaries} and {\em origin--destination matrices} data. The framework produces several validation metrics quantifying the {\em temporal}, {\em spatial} and {\em structural validity} of activity schedules generated by the model. These metrics can be used to assess the accuracy of the model, guide model development or compare the model accuracy to other models. We have applied VALFRAM to assess and compare the validity of three activity-based transport models of a real-world region comprising around 1~million inhabitants. In the test application, the framework correctly identified strong and weak aspects of each model, which confirmed the viability and usefulness of the framework. 
%\mjnote{pouzivam tady presnost, ale to jsme myslim jinde nedelali, priapadne nahrad validity}
%\todo{Nejaky future work}
% Six validation steps were carefully described to make their implementation as easy as possible. Each step was demonstrated on two models based on a real-world, central-european metropolitan area with population of approximately 1 million. The selection of the statistics was driven towards well-established methods which can be simply understood.
%\jdnote{future work: more complicated activity schedule structure with multimodal journeys?}
%\mcnote{do we even need to mention the future work?}

\begin{comment}
\subsection*{Acknowledgements}
This work was funded by Ministry of Education, Youth and Sports of Czech Republic (grants no. 7E12065 and LD12044), Technology Agency of the Czech Republic (grant no. TE01020155) and by the European Union Seventh Framework Programme FP7/2007-2013 (grant agreement no. 289067).
\end{comment}